\documentclass[12pt]{article}
\usepackage{amsmath,amssymb,euscript,array,cite,cancel,color}
\usepackage{cases,empheq}
\usepackage[colorlinks=true]{hyperref}
\usepackage{url}
\usepackage{bbm}

\usepackage[paper=a4paper,dvips,top=2.54cm,left=2.74cm,right=2.34cm,
    foot=1cm,bottom=2.54cm]{geometry}

\newcounter{multieqs}

\newcommand{\be}{\begin{equation}}
\newcommand{\ee}{\end{equation}}
\newcommand{\eq}[1]{(\ref{#1})}
\newcommand{\bit}{\begin{itemize}}  \newcommand{\eit}{\end{itemize}}

\def\a{\alpha}        
\def\b{\beta}         
\def\g{\gamma}  \def\G{\Gamma}    
\def\d{\delta}      
\def\e{\epsilon}          
          
\def\h{\eta}  
\def\k{\kappa}  
\def\l{{\lambda}} \def\L{\Lambda}  
\def\m{\mu} \def\n{\nu}  
\def\o{\omega}  
   
\def\r{\rho}  
\def\s{\sigma}    
  
\def\th{\theta}    
  
\def\z{\zeta}  

\def\na{\nabla}

  \def\cL{{\cal L}}  
\def\cM{{\cal M}} \def\cN{{\cal N}}

\def\cV{{\cal V}}

\def\d{\delta}  
  
\def\pa{\partial}   
  
\def\uno{\mbox{1 \kern-.59em {\rm l}}}

\newcommand{\tr}{\mbox{tr}}

\def\hlf{\frac{1}{2}}  
\def\ove#1{\frac{1}{#1}}

\def\bcomment#1{}

\newcommand{\lrbrk}[1]{\left(#1\right)}

\def\w{{\wedge}}

\author{Pichet Vanichchapongjaroen$~^a$\footnote{pichetv@nu.ac.th}
\\
\\
{\small $^a$ \it The Institute for Fundamental Study ``The Tah Poe Academia Institute'',}
\\
{\small\it Phitsanulok-Nakhonsawan Road,
Naresuan University, Phitsanulok 65000, Thailand}}

\title{Covariant M5--brane action with self--dual 3--form}

\begin{document}
\maketitle

\abstract{
In this work, we extend a theory which describes a linearly self--dual exact 3--form field
in six dimensional spacetime within the formalism of \href{https://arxiv.org/abs/1511.08220}{arXiv:1511.08220} and
\href{https://arxiv.org/abs/1903.12196}{arXiv:1903.12196}
%\cite{Sen:2015nph}, \cite{Sen:2019qit}
to an action which
fully
describes an M5--brane in the eleven--dimensional supergravity background.
The action we obtain has all the required symmetries including kappa--symmetry.
Although the derivation of the action and the proof the symmetries do not require one to express the action explicitly in terms of the original field contents,
we also give an initial attempt to give this expression.
}

\thispagestyle{empty}
\newpage
\tableofcontents

\setcounter{equation}0
\section{Introduction}
In order to construct a low--energy effective action for a single M5--brane,
it is important to base this on an
action which describes a chiral 2--form field, i.e. a 2--form field with self--dual 3--form field strength,
living in a six--dimensional spacetime.
This is because a chiral 2--form field is one of the field content
of M5--brane. Furthermore, it is
already not a straightforward matter to construct an action for a chiral 2--form field.

In the literature, there have been two main types chiral 2--form theories
in six dimensions
and the corresponding complete actions of single M5--brane
in eleven--dimensional supergravity background.
In one type, the actions are 
diffeomorhism invariant but not manifestly.
Chiral 2--form theories of this type are given in \cite{Henneaux:1988gg}, \cite{Perry:1996mk}, \cite{Chen:2010jgb}, \cite{Ko:2015zsy}.
Complete M5--brane actions of this type are constructed in \cite{Aganagic:1997zq}, \cite{Ko:2016cpw}.
In the other type, the actions are manifestly diffeomorphism invariant,
but auxiliary fields should be introduced.
One the of symmetries ensure that the auxiliary fields have no dynamics.
The approach to construct theories of this type is called the PST formalism. 
Actions for chiral 2--form of this type are given in \cite{Pasti:1995ii}, \cite{Pasti:1996va}, \cite{Pasti:1996vs}, \cite{Pasti:2009xc}.
Complete M5--brane actions of this type are constructed in
\cite{Pasti:1997gx}, \cite{Bandos:1997ui}, \cite{Ko:2013dka}, \cite{Ko:2017tgo}.
Actions for theories within either of these types
contains terms which describe dynamics of the chiral 2--form field. Self--duality condition for its 3--form field strength is obtained at the level of equations of motion. This is made possible due to a gauge symmetry usually called PST1 symmetry.

It could be possible to extend other kinds of chiral 2--form actions to obtain complete M5--brane actions. We find that one of the most promising recent constructions of a chiral 2--form theory is given in \cite{Sen:2015nph}, \cite{Sen:2019qit}. See also \cite{Lambert:2019diy} and \cite{Andriolo:2020ykk} for further developments.
The theory which is relevant to us describes
a 2--form field $P$ and a 3--form field $Q$ in a six--dimensional curved spacetime.
The 2--form field $P$ has the wrong sign of the kinetic term.
The 3--form field $Q$ is self--dual with respect to
flat metric. A certain combination of the 3--form field $Q$
with the curved metric gives rise to a 3--form field which is self--dual with respect to the curved metric.
At the level of equations of motion, the latter 3--form field is closed.
If the spacetime has a trivial topology, then this 3--form field is exact and hence,
since it is also self--dual, it is equivalent to a chiral 2--form field with self--dual 3--form field strength. Also at the level of equations of motion the 2--form field $P$ is decoupled.
Further investigation using Hamiltonian analysis also shows that the 2--form field $P$ is decoupled.
We will call the idea of this construction as Sen formalism. The analysis of the linear self--dual 3--form theory in Sen formalism has been quite throughout \cite{Sen:2015nph}, \cite{Sen:2019qit}, \cite{Lambert:2019diy}, \cite{Andriolo:2020ykk} which provides good background for us to seek to extend the theory.

The main goal of this paper is to extend the action in Sen formalism by nonlinearising the self--duality condition and extending the resulting action to fully
describe an M5--brane in the eleven--dimensional supergravity background.

Our strategy is to start by constructing an action for a nonlinear self--dual 3--form theory by using diffeomorphism invariance to fix the form of the action.
Then a 3--form source field can naturally be included in a similar way as in the linear self--dual case. One of gauge symmetries of this resulting theory has the gauge parameter which is subject to some condition. By including an appropriate 6--form source field,
it is possible to remove this restriction.
In fact, the 3--form and 6--form source fields can be realised as pull--backs of $C_3$ and $C_6$ fields from the eleven--dimensional supergravity background to the M5--brane worldvolume. Based on these results, a
complete action of an M5--brane in the eleven--dimensional supergravity background can be constructed. The action has all the required properties especially kappa--symmetry.

The rest of this paper is organised as follows.
In section \ref{sec:lincov}, we review the action of \cite{Sen:2015nph}, \cite{Sen:2019qit}. In particular, we focus on the prove of diffeomorphism invariance.
In section \ref{sec:nonlin}, we proceed to generalise to 
nonlinear self--dual case. In section \ref{sec:M5full}, we further generalise the action to a complete M5--brane action in eleven--dimensional supergravity background.
In the analysis explained so far, we do not need to know the exact expression of the action as a function of original field content. As long as all the properties are known, it is sufficient to prove the symmetry. Nevertheless, to prepare for other purpose, it might be useful to be able to explicitly express the action in terms of the original field content. In linear self--dual case, This task has been done \cite{Sen:2015nph}, \cite{Sen:2019qit}, \cite{Andriolo:2020ykk}. In section \ref{sec:solveR}, we will provide an initial attempt in nonlinear self--dual case.
Finally, in section \ref{sec:conclusion}, we give conclusions
and discuss possible directions for future works.

\setcounter{equation}0
\section{A covariant action for a linear self--dual field}\label{sec:lincov}
In this section, we review an action given in \cite{Sen:2015nph}, \cite{Sen:2019qit} which describes a self--dual 3--form field in six dimensions. In particular, we review basic properties which includes diffeomorphism invariance.

Consider a six--dimensional spacetime
with coordinates $x^\m, \m=0,1,\cdots,5$
and metric $g_{\m\n}.$
We define a differential $p-$form as
\be
\o_{(p)}
=\ove{p!}dx^{\m_1}\w\cdots dx^{\m_p}\o_{\m_p\cdots\m_1}.
\ee
Exterior derivatives and interior products
are defined to act from the right.
In Sen formalism,
there are two kinds
of Hodge star operator.
The first kind is the standard
Hodge star operator, which we label as $*.$
It acts as
\be
*dx^{\m_1}\w\cdots \w dx^{\m_p}
=\frac{(-1)^{p+1}}{(6-p)!\sqrt{-g}}dx^{\m_{p+1}}\w\cdots\w dx^{\m_{6}}g_{\m_{p+1}\n_{p+1}}\cdots g_{\m_6\n_6}\e^{\n_{p+1}\cdots\n_6\m_{1}\cdots\m_p},
\ee
where $\e^{\m_1\cdots\m_6}$ is a Levi--Civita symbol
defined such that $\e^{012\cdots5} = 1.$
The other kind which we label as $*'$
acts as
\be
*'dx^{\m_1}\w\cdots \w dx^{\m_p}
=\frac{(-1)^{p+1}}{(6-p)!}dx^{\m_{p+1}}\w\cdots\w dx^{\m_{6}}\h_{\m_{p+1}\n_{p+1}}\cdots \h_{\m_6\n_6}\e^{\n_{p+1}\cdots\n_6\m_{1}\cdots\m_p},
\ee
where $\h_{\m\n}$ is the flat metric.
In other words, $*'$ is defined with respect
to the flat metric.

Let us now review an action given in \cite{Sen:2015nph},
\cite{Sen:2019qit},
which is
\be\label{Sen-action}
S = \int \lrbrk{\hlf dP\w*' dP - 2Q\w dP + \cL_I(Q,g)},
\ee
where $P$ is a 2--form field, and $Q$ is a $*'$--self--dual 3--form field: $*'Q = Q.$
Let the variation of $\cL_I$ with respect to $Q$ be of the form
\be\label{dQcLI}
\d_Q \cL_I = 2\d Q\w R.
\ee
So $R$ is a $*'$--anti--self--dual 3--form: $*'R = -R.$
Let $H = Q-R$ be a $*$--self--dual 3--form:
\be\label{H-selfdual}
*H = H.
\ee
The condition \eq{H-selfdual} can be thought of as an extra condition
on $R.$ It is shown by construction \cite{Sen:2015nph},
\cite{Sen:2019qit}, \cite{Andriolo:2020ykk} that
this condition along with $*'$--anti--self--duality of $R$
can be used to completely determine $R.$

Varying the action \eq{Sen-action} gives
\be
\d S
= \int \lrbrk{d\d P\w(-dP + *' dP + 2Q) + 2\d Q\w \lrbrk{R - \hlf dP + \hlf *' dP}}.
\ee
Therefore, equations of motion are
\be\label{QP}
d(-dP + *' dP + 2Q) = 0,
\ee
\be\label{RP}
R = \hlf dP - \hlf *' dP.
\ee
Eliminating $dP$ from eq.\eq{RP}
and substituting into eq.\eq{QP}
gives
\be\label{dH0}
dH = 0.
\ee
In spacetime with trivial topology, $H$ is exact.
It can be seen from eq.\eq{H-selfdual}
and eq.\eq{dH0}
that, in spacetime with trivial topology,
combining the metric 
$g_{\m\n}$ and the field $Q$ in a certain way
gives rise to a field $H$ which is
$*$--self--dual off--shell
and is exact on--shell.
The field $P,$ which is unphysical due to the wrong sign of its kinetic term, is decoupled.
In particular, it has been shown \cite{Sen:2019qit}
that a combination
\be
H_{(s)}
=\hlf(dP + *' dP) + Q,
\ee
is decoupled
and is invariant under a diffeomorphism
and gauge transformations.
Furthermore, after extending the theory
to an abelian $(2,0)$ theory,
$H_{(s)}$ is a singlet under supersymmetry transformation
\cite{Lambert:2019diy}, \cite{Andriolo:2020ykk}.

From the review above it can be seen that
the field content $P$ and $Q$
do not have standard properties.
In fact, as we will also review later,
these fields do not transform as standard differential forms
under diffeomorphism transformations.
Hence, they are called pseudoforms.

It is interesting to compare Sen formalism with PST formalism
in the construction of an action for a 2--form field with self--dual 3--form field strength.
In short, Sen formalism gives a 3--form field which is self--dual off--shell and is closed on--shell. In a spacetime with trivial topology, this field is exact.
On the other hand, the construction of PST formalism describes a 2--form field. So its 3--form field strength is exact off--shell.
The self--duality of 3--form field strength is obtained on--shell. The analysis can be done in the case of spacetime with trivial topology as well as spacetime with nontrivial topology \cite{Bandos:2014bva}, \cite{Isono:2014bsa}.

More detailed discussions are as follows.
We have seen that in Sen formalism, the action \eq{Sen-action} of the 
pseudoforms $P$ and $Q = *'Q$ is constructed such that the form of $\cL_I$ allows a certain combination of
the metric and $Q$ to give rise to $H$ which is $*$--self--dual. At the level of the equation of motion, the field $H$ is closed
and the field $P$ is decoupled.
If the spacetime has trivial topology,
the field $H$ is then exact on--shell.
In the PST formalism, the 2--form field $B$ is introduced
whose field strength will be self--dual at the level of equations of motion. In order to achieve this while making the action manifestly diffeomorphism invariant, one or more auxiliary scalar fields
are introduced. In this formalism, apart from the standard tensor gauge symmetry, there are two important gauge symmetries called PST1 and PST2 symmetries. The PST1 symmetry is used to ensure that the equation of motion of $B$ reduces to self--duality condition for $dB.$

In Sen formalism, the presence of a non--standard Hodge star operator $*'$ suggests that the action \eq{Sen-action} is not manifestly diffeomorphism invariant. In \cite{Sen:2015nph}, \cite{Sen:2019qit}, \cite{Andriolo:2020ykk}, the expression of 
$\cL_I$ is given at the start. Then the non--standard diffeomorphism transformation of $P$ and $Q$ which leaves the action invariant is determined.
For us, in order to gain some insights as a preparation for nonlinearisation,
let us proceed in the opposite direction.
We will simply impose diffeomorphism transformations
on $P$ and $Q$ as given in \cite{Sen:2015nph}, \cite{Sen:2019qit}, \cite{Andriolo:2020ykk}, then proceed to determine the form of $\cL_I.$

Let $P$ and $Q$ transform under infinitesimal diffeomorphism $x^\m\mapsto x^\m + \xi^\m$
as
\be\label{PQdiffeo}
\d_\xi P = i_\xi H,\qquad
\d_\xi Q = -\hlf (1+*')d\d_\xi P.
\ee
Diffeomorphism variation on the action \eq{Sen-action}
is then
\be\label{dxiS}
\d_\xi S
=\int \lrbrk{-2 H\w di_\xi H + \frac{\d\cL_I}{\d g_{\m\n}}\d_\xi g_{\m\n}}.
\ee
Consider
\be\label{HdiH-curved-lin}
H\w di_\xi H
=-\hlf d^6x\sqrt{-g}H^{\m\n\l}\na_\l(\xi^\r H_{\m\n\r}).
\ee
Using Leibniz rule, we obtain
\be\label{HdiH-curved-lin-1}
H\w di_\xi H
=\textrm{tot.} + \hlf d^6 x\sqrt{-g}\xi^\r H_{\m\n\r}\na_\l H^{\m\n\l},
\ee
where $\textrm{tot.}$ are total derivative terms.
On the other hand, by using self--duality condition,
eq.\eq{HdiH-curved-lin} becomes
\be\label{HdiH-curved-lin-2}
H\w di_\xi H
=\textrm{tot.}
-\hlf d^6 x\sqrt{-g}H^{\m\n\l}\na_\l \xi^\r H_{\m\n\r}
- \hlf d^6 x\sqrt{-g}\xi^\r H_{\m\n\r}\na_\l H^{\m\n\l}.
\ee
Therefore, adding eq.\eq{HdiH-curved-lin-1} and eq.\eq{HdiH-curved-lin-2}
together gives
\be
2H\w di_\xi H
=\textrm{tot.}
-\hlf d^6 x\sqrt{-g}H^{\m\n\l}\na_\l \xi^\r H_{\m\n\r}.
\ee
So the action is diffeomorphism invariant
if the condition
\be\label{dcLI}
\frac{\d\cL_I}{\d g_{\m\n}}\d_\xi g_{\m\n}
=-\hlf d^6 x\sqrt{-g}H^{\m\n\l}\na_\l \xi^\r H_{\m\n\r}
\ee
is satisfied.

It is possible to rewrite RHS of eq.\eq{dcLI}.
For this, consider
\be\label{dxi-star-H}
\begin{split}
(\d_\xi *)H
&= \d_\xi (*H) - *\d_\xi H\\
&= (1-*)\d_\xi H.
\end{split}
\ee
Direct calculation using
$\d_\xi H = -\cL_\xi H$
gives
\be\label{dxi-star-H-1}
\begin{split}
(\d_\xi *)H
&=\na_\m\xi^\m H + \hlf dx^{\m}\w dx^{\n}\w dx^{\r}(\na^\l\xi_\m + \na_\m\xi^\l)H_{\l\n\r}\\
&=(1-*)\hlf dx^{\m}\w dx^{\n}\w dx^{\r}\na_\m\xi^\l H_{\l\n\r}.
\end{split}
\ee
On the other hand, using $H = Q-R$
on eq.\eq{dxi-star-H} gives
\be\label{dxi-star-H-2}
(\d_\xi *)H
=
-(1-*)\frac{\d R}{\d g_{\l\s}}\d_\xi g_{\l\s}.
\ee
Comparing eq.\eq{dxi-star-H-1} and eq.\eq{dxi-star-H-2}
gives
\be
-(1-*)\frac{\d R}{\d g_{\l\s}}\d_\xi g_{\l\s}
=(1-*)\hlf dx^{\m}\w dx^{\n}\w dx^{\r}\na_\m\xi^\l H_{\l\n\r}.
\ee
Applying wedge product with $H$ to the above equation and using eq.\eq{dcLI}
gives
\be
\frac{\d\cL_I}{\d g_{\m\n}}\d_\xi g_{\m\n}
=Q\w\frac{\d R}{\d g_{\m\n}}\d_\xi g_{\m\n},
\ee
which can be integrated to give
\be
\cL_I = Q\w R + \z(Q).
\ee
By taking derivative with respect to $Q$
and using $H\w\d_Q H =  0,$ we obtain $\z(Q) = 0,$
and hence
\be
\cL_I = Q\w R,
\ee
which is as required.

\setcounter{equation}0
\section{A covariant action for a nonlinear self-dual field}\label{sec:nonlin}

In this section, we present a construction of a covariant action for a nonlinear self--dual field.
Apart from the fields $P, Q,$ and $g,$
let us also introduce a 3--form source.
The action should take the form
\be\label{Snonlin-0}
S = \int \lrbrk{\hlf dP\w*' dP - 2Q\w dP + \cL_I(Q,g,J)},
\ee
where $P$ and $Q$ are as defined in the previous section,
and $J$ is a 3--form source.
Let us write the variation of $\cL_I$ with respect to $Q$
as
\be\label{dQcLI-nonlin}
\d_Q\cL_I 
=2\d Q\w R(Q,g,J).
\ee                
So $R$ is $*'$--anti--self--dual.
The equation of motion is modified from eq.\eq{dH0} to
\be                
dH^J = dJ,         
\ee
where $H^J = Q-R+J.$ Furthermore, as in the linear self--dual case, the combination $H_{(s)}$ is also decoupled.
We require $H^J$ to be 
nonlinear $*$--self--dual,
that is
\be\label{nonlin-self-dual-cond}
*H^J = \cV(H^J,g).
\ee
Again, eq.\eq{nonlin-self-dual-cond} can be thought of as a condition given to fix the form of $\cL_I.$
It is not an on--shell condition.

For the purpose of this paper, whose main goal is to construct an M5--brane action, it is sufficient to consider $\cV$ of the form
\be\label{cV-nonlin}
\cV = F_1([(K^J)^2]) H^J + F_2([(K^J)^2]) (H^J)^3,
\ee
where
\be
((H^J)^3)_{\m\n\r} \equiv H^J_{[\m|\n'\r'}(H^J)^{\m'\n'\r'}H^J_{\m'|\n\r]},\qquad
(K^J)_\m{}^\l
\equiv H^J_{\m\n\r}(H^J)^{\l\n\r},
\ee
and the notation $[\cdot]$ stands for trace over spacetime indices.
It is sufficient to allow 
$F_1$ and $F_2$ to only depend on $[(K^J)^2].$ This is because $[K^J]$ is expressible as a function of $[(K^J)^2]$ by using
\be\label{trKintrK2}
[K^J] = -\frac{F_2}{F_1}[(K^J)^2],
\ee
which can be obtained by considering
$(H^J)^{\m\n\r}\cV_{\m\n\r}$ and using
eq.\eq{nonlin-self-dual-cond} and eq.\eq{cV-nonlin}.
As to be given in details later, eq.\eq{nonlin-self-dual-cond}--\eq{cV-nonlin} with a specific choice of $F_1$ and $F_2$ describes nonlinear $*$--self--duality condition for M5--brane.

\subsection{Useful identities}
Before proceeding to derive the action, it would be useful to first present some identities from nonlinear self--duality conditions.

It can be seen that eq.\eq{nonlin-self-dual-cond}
with $\cV$ given by eq.\eq{cV-nonlin} satisfies
\be\label{cond-nonlin-1}
\cV^{\m\n\l}H^J_{\m\n\r}
=(H^J)^{\m\n\l}\cV_{\m\n\r},
\ee
and
\be\label{cond-nonlin-3}
(H^J)^{\m\n\r}\d\cV_{\m\n\r}
=\cV^{\m\n\r}\d (H^J)_{\m\n\r}
+\d U,
\ee
where
\be\label{UfromF1F2}
U = \int d [(K^J)^2]\lrbrk{\frac{F_2}{2} - \frac{[(K^J)^2] F_2 F_1'}{F_1} + [(K^J)^2] F_2'},
\ee
and
\be\label{dingHJ}
\d = \frac{\d}{\d g_{\m\n}}\d g_{\m\n} + \frac{\d}{\d (H^J)_{\m\n\r}}\d (H^J)_{\m\n\r}.
\ee
Note that these conditions can alternatively be written as
\be
\e^{\m\n\l\m'\n'\r'}H^J_{\m'\n'\r'}H^J_{\m\n\r}
=\e^{\m\n\l\m'\n'\r'}\cV_{\m'\n'\r'}\cV_{\m\n\r},
\ee
\be
\e^{\m\n\r\m'\n'\r'}H^J_{\m\n\r}\d H^J_{\m'\n'\r'}
=\e^{\m\n\r\m'\n'\r'}\cV_{\m\n\r}\d \cV_{\m'\n'\r'}
+6\sqrt{-g}\d U.
\ee
Alternatively, the condition
\eq{cond-nonlin-3} can be expressed using differential forms as
\be\label{cond-nonlin-3-2}
\cV\w\d\cV - H^J\w\d H^J = -\ove{3!}d^6 x\sqrt{-g}\d U.
\ee
Note also that there is an identity
\be\label{cVdH-2}
\cV^{\m\n\r}\d H^J_{\m\n\r} = -\hlf\frac{\d U}{\d H^J_{\m\n\r}}\d H^J_{\m\n\r}.
\ee

Note that these identities are derived based on the form of $\cV$ as given in eq.\eq{cV-nonlin}.
This form alone, however, is not enough to ensure that eq.\eq{nonlin-self-dual-cond} is consistent.
There should be extra restrictions to ensure this.
By counting the number of components, eq.\eq{nonlin-self-dual-cond} is a system of $20$ equations. 
In order for it to truly be a nonlinear self--duality condition, the system should contain $10$ independent equations. This requirement gives consistency conditions which would fix the form of $\cV.$
Note, however, that we do not yet need to impose these extra conditions to derive the action. So we postpone the discussion of the consistency conditions to subsection \ref{subsec:nonlinsd}.

\subsection{Derivation of the action using diffeomorphism}

Let us now discuss the derivation of the action.
We are going to do this by demanding that the action is diffeomorphism invariant.
The consideration is in parallel to the linear self--dual case.
Let $P, Q$ transform as
\be\label{PQdiffeo-nonlin}
\d_\xi P = i_\xi (H^J - J),\qquad
\d_\xi Q = -\hlf (1+*')d\d_\xi P.
\ee
The action then transforms as
\be\label{dS-diffeo-nonlin}
\begin{split}
\d_\xi S
&=\int\lrbrk{-2(H^J-J)\w di_\xi(H^J - J) + \frac{\d\cL_I}{\d g_{\m\n}}\d_\xi g_{\m\n} + \frac{\d\cL_I}{\d J_{\m\n\r}}\d_\xi J_{\m\n\r}}\\
&=\int\lrbrk{H^J\w\d_\xi H^J - 2H^J\w\d_\xi J + J\w\d_\xi J
+ \frac{\d\cL_I}{\d g_{\m\n}}\d_\xi g_{\m\n} + \frac{\d\cL_I}{\d J_{\m\n\r}}\d_\xi J_{\m\n\r}},
\end{split}
\ee
where we used $\d_\xi H^J = -\cL_\xi H^J$ and $\d_\xi J = -\cL_\xi J.$
Direct calculation using $\d_\xi H^J = -\cL_\xi H^J$ gives
\be\label{HJdHJ}
\begin{split}
H^J\w\d_\xi H^J
&=\ove{3!}d^6 x\sqrt{-g}\cV^{\m\n\r}\xi^\l\na_\l H^J_{\m\n\r}
+\hlf d^6 x\sqrt{-g}\cV^{\m\n\r}\na_\m \xi^\l H^J_{\l\n\r}.
\end{split}
\ee
Imposing
eq.\eq{cond-nonlin-1}--\eq{cond-nonlin-3}
on eq.\eq{HJdHJ},
we obtain
\be\label{iden-nonlin-1}
H^J\w\d_\xi H^J
=\textrm{tot.}
-\ove{4} d^6x\sqrt{-g}\cV^{\m\n\l}\d_\xi g_{\r\l} (H^J)_{\m\n}{}^\r
-\ove{24}d^6x \sqrt{-g} U \d_\xi g_{\m\n}g^{\m\n}.
\ee
So after substituting into eq.\eq{dS-diffeo-nonlin}, we see that there are two types of expressions. The first type contains terms proportional to $\d_\xi g_{\m\n}.$ The second type contains  terms proportional to $\d_\xi J_{\m\n\r}.$ Let us suppose that these two types should vanish separately. So as part of this, we should set
\be
\frac{\d\cL_I}{\d g_{\m\n}}
=\ove{4} d^6x\sqrt{-g}\cV^{\l\s\m} (H^J)_{\l\s}{}^\n
+\ove{24}d^6x \sqrt{-g} Ug^{\m\n}.
\ee
It is possible to simplify the first term on RHS.
For this, let us use the condition \eq{cond-nonlin-3-2} and the identity
\be
\cV\w\d \cV + H^J\w\d H^J = -\hlf\cV^{\l\s\m} H^J_{\l\s}{}^\n\sqrt{-g}\d g_{\m\n}.
\ee
This gives
\be
\lrbrk{\frac{\d\cL_I}{\d g_{\m\n}}}_{Q,J}
=\ove{12} d^6x\sqrt{-g}\lrbrk{\frac{\d U}{\d g_{\m\n}}}_{H^J}
+\ove{24}d^6x \sqrt{-g} U g^{\m\n}.
\ee
By using eq.\eq{cond-nonlin-3-2} and eq.\eq{iden-nonlin-1}, we obtain
\be
d^6 x\lrbrk{\frac{\d U}{\d g_{\m\n}}}_{Q,J}
=d^6 x\lrbrk{\frac{\d U}{\d g_{\m\n}}}_{H^J}
+\frac{12}{\sqrt{-g}}H^J\w\lrbrk{\frac{\d H^J}{\d g_{\m\n}}}_{Q,J},
\ee
and hence
\be\label{dgcLI}
\begin{split}
\lrbrk{\frac{\d\cL_I}{\d g_{\m\n}}}_{Q,J}
&=
\ove{24}d^6x \sqrt{-g} U g^{\m\n}
+
\ove{12} d^6x\sqrt{-g}\lrbrk{\frac{\d U}{\d g_{\m\n}}}_{Q,J}
- H^J\w\lrbrk{\frac{\d H^J}{\d g_{\m\n}}}_{Q,J}\\
&=
\ove{24}d^6x \sqrt{-g} U g^{\m\n}
+
\ove{12} d^6x\sqrt{-g}\lrbrk{\frac{\d U}{\d g_{\m\n}}}_{Q,J}
+(Q+J)\w\lrbrk{\frac{\d R}{\d g_{\m\n}}}_{Q,J}.
\end{split}
\ee
Integrating gives
\be
\cL_I
=\ove{12}d^6 x\sqrt{-g}U + (Q+J)\w R +\z(Q,J).
\ee
Taking derivative with respect to $J$ gives
\be
\begin{split}
\d_J\cL_I
&=\ove{12}d^6 x\sqrt{-g}\d_J U + \d_J J\w R
+ (Q+J)\w \d_J R + \d_J\z\\
&=H^J\w\d_J H^J + \d_J J\w R
+ (Q+J)\w \d_J R + \d_J\z.
\end{split}
\ee
On the other hand, demanding terms proportional to $\d_\xi J_{\m\n\r}$ in eq.\eq{dS-diffeo-nonlin} to vanish gives
\be\label{dJcLI}
\d_J\cL_I = 2H^J\w\d_J J - J\w\d_J J.
\ee
Comparing the two expressions of $\d_J\cL_I$ gives
\be
\d_J\z
=Q\w\d_J J,
\ee
and hence
\be
\z = Q\w J + \bar{\z}(Q).
\ee
So
\be
\cL_I
=\ove{12}d^6 x\sqrt{-g}U + (Q+J)\w R + Q\w J + \bar{\z}(Q).
\ee
Taking derivative with respect to $Q$ gives
\be
\begin{split}
\d_Q\cL_I
&=H^J\w\d_Q H^J
+\d_Q Q\w R + (Q+J)\w\d_Q R
+\d_Q Q\w J + \d_Q\bar{\z}\\
&=2\d_Q Q\w R+ \d_Q\bar{\z},
\end{split}
\ee
and hence
\be
\d_Q\bar{\z} = 0.
\ee

So from the consideration of diffeomorphism invariance, we obtain the action
\be\label{nonlinSenAction}
S = \int \lrbrk{\hlf dP\w*' dP - 2Q\w dP
+\ove{12}d^6 x\sqrt{-g}U + (Q+J)\w R + Q\w J},
\ee
where $U$ is a function of $H^J(Q,J,g)$ and $g.$
The form of $U(H^J(Q,J,g),g)$ and $R(Q,J,g)$ 
should be determined from $*$--self--duality condition eq.\eq{nonlin-self-dual-cond}
provided that it is consistent with $*'$--self--duality of $Q$
and that the conditions \eq{cond-nonlin-1} and \eq{cond-nonlin-3}
are satisfied.
Note that in the linear $*$--self--dual case, the conditions \eq{cond-nonlin-1} and \eq{cond-nonlin-3}
are satisfied with $U=0.$
In this case, the action \eq{nonlinSenAction} also reduces to the form given in \cite{Sen:2015nph}, \cite{Sen:2019qit}, \cite{Andriolo:2020ykk}.

It can be seen that just like the linear self--dual case, the field $P$ is decoupled at the level of equations of motion. In the nonlinear self--dual case, the decoupling of $P$ can also be seen from Hamiltonian analysis \cite{Sen:2019qit}, \cite{Andriolo:2020ykk}. In fact, the result of the analysis of \cite{Sen:2019qit} holds for actions of the form \eq{Snonlin-0}. Since our nonlinear action falls in this case, the result of Hamiltonian analysis of \cite{Sen:2019qit} also valid for our case.
Although the Hamiltonian analysis of 
\cite{Sen:2019qit}, \cite{Andriolo:2020ykk} treat only $P$ and $Q$ in the canonical formulation,
it can be expected that even if every field is treated in the canonical formulation the field $P$ still decouples from the theory as long as it does not couple to the metric and $J.$ This is indeed the case for the action \eq{nonlinSenAction}.

In order to gain further insights on the action \eq{nonlinSenAction},
it would be useful to consider the case of flat spacetime\footnote{We thank Dmitri Sorokin for the correspondence on this issue}.
In this case, the two types of metrics $g_{\m\n}$ and $\h_{\m\n}$
should coincide and hence
one would expect that the fields $P$ and $Q$ should transform
in the standard way under Lorentz transformation.
It turns out that this standard Lorentz transformation
is not obtainable from the nonstandard diffeomorphism transformation
\eq{PQdiffeo}.
In fact, the theory \eq{nonlinSenAction}
in flat spacetime has
two types of Lorentz symmetries:
the standard one and the nonstandard one which
is obtained from 
diffeomorphism transformation
\eq{PQdiffeo} with $\xi_\m = \o_{[\m\n]}x^\n.$
In order for the action to possess
the nonstandard Lorentz symmetry,
the conditions \eq{cond-nonlin-1} and \eq{cond-nonlin-3}
should still be satisfied.
When promoted to curved spacetime,
it turns out that only the nonstandard Lorentz symmetry
survives and is indeed promoted to the nonstandard
diffeomorphism transformation \eq{PQdiffeo}.

\subsection{Symmetries}
Apart from being invariant under diffeomorphism by construction, the action \eq{nonlinSenAction} also enjoys the gauge symmetries as in the case of linear $*$--self--dual case \cite{Sen:2019qit}, \cite{Andriolo:2020ykk}.
The invariance under $P\to P+d\l$ is straightforward. To check another gauge symmetry, let us note that subject to
\be\label{PQtrafo}
\d Q = -\hlf(1+*')d\d P,
\ee
the variation of the action is given by
\be\label{deltaS}
\begin{split}
\d S
&=\int\bigg(-2(H^J - J)\w d\d P
+(2H^J- J)\w\d J\\
&\qquad\qquad+\hlf\ove{3!}d^6x\d\sqrt{-g}U
+\ove{4}d^6x\sqrt{-g}\cV^{\l\s\m}(H^J)_{\l\s}{}^\n\d g_{\m\n}
\bigg).
\end{split}
\ee
So the action is obviously invariant under gauge transformation
\be\label{C3variation}
\d P = \L,\qquad
\d J = d\L,\qquad
\d g_{\m\n} = 0,
\ee
where $\L$ is subject to
\be
\int\L \w dJ = 0.
\ee
In fact, as shall be seen in section \ref{sec:M5full},
a 6--form source field with appropriate transformation
can be included into the action so that the resulting action
is gauge invariant without the need to impose extra condition.

By considering eq.\eq{deltaS}, it can be seen that energy--momentum tensor can easily be derived
and can be shown to be conserved.
The analysis is similar to the case of linear self--duality given in \cite{Andriolo:2020ykk}.
In the case of nonlinear self--duality, energy--momentum tensor is given by
\be
\begin{split}
T_{\m\n}
&=\frac{-2}{\sqrt{-g}}\frac{\d S}{\d g^{\m\n}}\\
&=\hlf\ove{3!} U g_{\m\n}
+\hlf\cV^{\l\s}{}_{(\m}H^J_{\n)\l\s}.
\end{split}
\ee
The conservation of the energy--momentum tensor
can also be shown by using
the diffeomorphism invariance of the action.
With the help of eq.\eq{deltaS}, we obtain
\be
0=\d_\xi S
=\int(-2H^J\w i_\xi dH^J + d^6x\sqrt{-g}\na_\m T^{\m\n}\xi_\n).
\ee
Imposing equation of motion $dH^J = dJ,$
we obtain
\be
\na_\m T^{\m\n}
=-4\cV_{\r\s\l}\na^{[\n}J^{\r\s\l]}.
\ee
So when there is no source,
the energy--momentum tensor 
is conserved.

In the case of linear
$*$--self--duality, in which $U=0,$
the energy--momentum tensor reduces to the one given in \cite{Andriolo:2020ykk}.
The conservation then also follows.

\setcounter{equation}0
\section{A covariant M5--brane action in Green-Schwarz formalism}\label{sec:M5full}

In this section, we are going to present an extension of the action \eq{nonlinSenAction} with an appropriate $U$ to describe a complete M5--brane action embedded in the eleven--dimensional supergravity background.

\subsection{Nonlinear $*$--self--duality condition from M5--brane}
Let us start by showing that
the bosonic chiral 2--form field which is one of the fields
on M5--brane worldvolume can be described by
the action \eq{nonlinSenAction} with appropriate $U,$
to be given below.
Our strategy is to first obtain the form of eq.\eq{nonlin-self-dual-cond}
in flat six-dimensional spacetime and without source.
Since this equation should be Lorentz covariant,
the subsequent introduction of the curved metric and source should be straightforward.

One way to obtain nonlinear self--duality condition
for the 3--form field strength of the chiral 2--form
is by considering \cite{Howe:1997vn},
which derives the condition by using superembedding formalism.
In this description, there are two kinds of 3--form fields.
One of them is $h,$ which is linear self--dual
but is not a field strength of any 2--form field.
Another is $H,$ which is a field strength of some 2--form field,
but $H$ itself is not linearly self--dual.
Instead, it satisfies a
nonlinear self--dual condition which can be obtained by solving
\be\label{Hh}
\ove{4}H_{\m\n\r} = (m^{-1})_\m{}^\l h_{\l\n\r},
\ee
where
\be
m_\m{}^\n = \d_\m^\n - 2 k_\m^\n,\qquad
k_\m^\n = h_{\m\r\s}h^{\n\r\s}.
\ee

In order to obtain the self-duality condition in the form
\eq{nonlin-self-dual-cond},
the field $h_{\m\n\r}$ should be eliminated.
One way to do this is to first make a $5+1$ split of spacetime indices \cite{Howe:1997vn} to obtain $\cV_{\m\n 5}$ in terms of $H_{\m\n 5}.$
One may then try to propose an ansatz for the covariant form and demand that it is consistent with the form obtained from a $5+1$ split.
Another way to do this is by covariantly eliminating the field $h_{\m\n\r}$ \cite{Cederwall:1997gg}, \cite{Sezgin:1998tm}.

Having been able to put eq.\eq{Hh} in the form of eq.\eq{nonlin-self-dual-cond}, one can then include the curved metric and the source. This finally gives
\be\label{cV-M5}
\cV_{\m\n\r}
=\frac{2+W}{\sqrt{2}\sqrt{6+W}}H^J_{\m\n\r} - \ove{\sqrt{2}\sqrt{6+W}}((H^J)^3)_{\m\n\r},
\ee
where
\be
W = \sqrt{4+[(K^J)^2]/3}.
\ee
By using eq.\eq{trKintrK2} and eq.\eq{UfromF1F2}, we obtain
\be\label{trKastrK2}
[K^J] = -6 + \sqrt{3}\sqrt{12+[(K^J)^2]},
\ee
and
\be
U = -24\sqrt{1 + \frac{[K^J]}{24}}\equiv U_{\textrm{M}5}(H^J,g).
\ee
Alternatively, we may reexpress $U_{\textrm{M}5}(H^J,g)$
to the form introduced in \cite{Sezgin:1998tm}. In this case, eq.\eq{cVdH-2} is given by
\be
\cV^{\m\n\r}\d(H^J)_{\m\n\r}
=12\d(H^J)_{\m\n\r}\frac{\d}{\d (H^J)_{\m\n\r}}\sqrt{1+\ove{12}[K^J] + \ove{288}[K^J]^2 - \ove{96}[(K^J)^2]}.
\ee

In fact, we also need to ensure that the analysis so far in this section
is consistent after expressing $H$ in terms of $Q.$
We will discuss in subsection \ref{subsec:nonlinsd} that this is indeed the case.

\subsection{Extension to a complete M5--brane action}
The action \eq{nonlinSenAction} with $U=U_{\textrm{M}5}$ can be promoted to a full M5--brane action in Green--Schwarz formalism. For this, let us first state a convention which we use to describe the target superspace.

The eleven-dimensional target superspace is parametrised by $Z^{\cal M} = (X^M,\th),$ where $X^M$ are eleven bosonic coordinates and $\th$ are $32$ real fermionic coordinates.
The supervielbeins are
$E^A(Z) = dZ^{\cM} E_{\cM}{}^A(Z)$
and $E^\a(Z) = dZ^{\cM}E_{\cM}{}^\a(Z)$
where $A = 0,1,\cdots,10$ is a bosonic tangent space indices and $\a = 1,2,\cdots,32$ are fermionic tangent space indices. The kappa-symmetry would require that the superspace torsion has to satisfy a constraint
\be
T^A
=-iE^\a\G^A_{\a\b}E^\b.
\ee
The target superspace also contains a 3--form gauge superfield
\be
C_3(Z) = \ove{3!}dZ^{\cM_1} dZ^{\cM_2} dZ^{\cM_3}C_{\cM_3\cM_2\cM_1}(Z),
\ee
and its $C_6(Z)$ dual.
Kappa--symmetry would also require the field strengths of $C_3(Z)$ and $C_6(Z)$ to satisfy the following constraints
\be \label{F47}
\begin{split}
dC_3
&= -\frac i2 E^AE^BE^\alpha E^\beta(\Gamma_{BA})_{\alpha\beta}
+\cdots\,,\\
dC_6-C_3dC_3
&= \frac{2i}{5!} E^{A_1}\cdots E^{A_5}E^\alpha E^\beta(\Gamma_{A_5\cdots A_1})_{\alpha\beta}
+\cdots\,,
\end{split}
\ee
where $\cdots$ are terms which are not relevant for kappa--symmetry calculation.

The embedding of an M5--brane worldvolume into the target superspace
is described by the embedding functions $Z^\cM(x).$
The induced metric on the M5--brane worldvolume is given by
\be
g_{\m\n}(x) = E_\m^A(x) E_\n^B(x)\h_{AB},
\ee
where
\be
E_\m^A(x) \equiv \pa_\m Z^{\cN}E_{\cN}{}^A(Z(x)).
\ee
The M5--brane worldvolume still contains the fields $P(x)$ and $Q(x).$ As for source field $J(x),$ it is realised as the pull--back of $C_3$ to the M5--brane worldvolume. So let us denote $J$ as $C_3$
and
redefine the notation of $H^J$ from the previous section to $F.$ That is
\be
F = H + C_3,\qquad H = Q-R.
\ee

As a first step to promote the action \eq{nonlinSenAction},
we need to include a Wess--Zumino term.
In fact, it could be expected that
only $C_6$ term can be added into the action.
An inclusion of $F\w C_3$ which is the other part of the Wess--Zumino term would modify the equation of motion of $Q$
which is already in the desirable form.
Correspondingly, the need to add only $C_6$ term
can be seen from a gauge symmetry requirement.
If there is no condition imposed on the gauge parameter,
the action \eq{nonlinSenAction} would not be invariant under
the transformation \eq{C3variation}.
The transformation on the extra term $C_6$ cancels out the non--zero
contribution from the variation of the action \eq{nonlinSenAction}.
The resulting action is then given by
\be\label{Sfull}
S_{\textrm{M}5} = \int \lrbrk{\hlf dP\w*' dP - 2Q\w dP
+\ove{12}d^6 x\sqrt{-g}U_{\textrm{M}5}(F,g) + Q\w R + C_6 + F\w C_3}.
\ee
Note that the last two terms are combined into Wess--Zumino term. Note also that the field $F$ appearing in the action is actually a function of $Q, C_3, g.$
The action \eq{Sfull}
is invariant under the gauge transformation
which is promoted from eq.\eq{C3variation},
that is
\be
\begin{split}
\d P = \L_2,\qquad
\d Q = -\hlf(1+*')d\d P,\qquad
\d C_3 = d\L_2,\\
\d C_6 = d\L_5 + d\L_2\w C_3,\qquad
\d g_{\m\n} = 0,
\end{split}
\ee
where the gauge transformations on $C_3$ and $C_6$
are such that their field strengths are gauge invariant.

The action \eq{Sfull} should be invariant under the local fermionic kappa--symmetry
with parameters $\k^\a(x)$ which transforms the pull--back of the embedding functions such that
\be
i_\k E^\a
\equiv \d_\k Z^\cM E^\a_\cM
=\hlf(1+\bar{\G})^\a{}_\b\k^\b,\qquad
i_\k E^A
\equiv \d_\k Z^\cM E^A_\cM
=0.
\ee
The induced metric then transforms as
\be
\d_\k g_{\m\n} = -4i E_{(\m}^\a(\G_{\n)})_{\a\b}i_\k E^\b.
\ee
Up to this point, when considering $\d_\k S_{\textrm{M}5},$
there are still unknown quantities
which are $\d_\k P, \d_\k Q,$ and $di_\k C_3.$
It turns out that by imposing
\be
\d_\k P = i_\k C_3,\qquad
\d_\k Q = -\hlf(1+*')d\d_\k P,
\ee
the expression $di_\k C_3$ no longer appears in $\d_\k S_{\textrm{M}5}.$
We then see that the action transforms as
\be
\d_\k S_{\textrm{M}5}
=-2i\int d^6 x\sqrt{-g}\bar{E}_\m^\a\lrbrk{Y^\m}_{\a\b}(1+\bar{\G})^\b{}_\g\k^\g,
\ee
where
\be
Y^\m\equiv\G^\m\bar\g - \hlf\tilde{F}^{\m\n\r}\G_{\n\r} +\ove{4}\tilde{F}^{\r\s(\m} F_{\r\s}{}^{\n)}\G_\n -\sqrt{1+\frac{F_{\s\n\r}F^{\s\n\r}}{24}}\G^\m,
\ee
with
\be
\G_\m\equiv E_\m^A\G_A,\qquad
\bar\g\equiv \ove{6!\sqrt{-g}}\e^{\m_1\cdots\m_6}\G_{\m_1\cdots\m_6}.
\ee
The quantity $(1+\bar\G)/2$ is a projector of rank $16$
with \cite{Cederwall:1997gg}, \cite{Sezgin:1998tm}
\be
\sqrt{1+\frac{F_{\m\n\r}F^{\m\n\r}}{24}}\,\bar\G
=\bar\g - \ove{12}\tilde{F}^{\m\n\r}\G_{\m\n\r}.
\ee
So that
\be
\bar\G^2 = 1,\qquad
\tr\bar\G = 0.
\ee

It would be useful to compare the action \eq{Sfull} with M5--brane action from other formalisms.
In particular, let us compare on--shell actions.
The action \eq{Sfull} is given on--shell as
\be\label{Sfull-onshell}
S_{\textrm{M}5}^{\textrm{on--shell}} = -2\int
d^6 x\sqrt{-g}\sqrt{1+\frac{F_{\m\n\r}F^{\m\n\r}}{24}} + \int(C_6+F\w C_3).
\ee

An action which is closely related to our construction is the one presented in \cite{Sezgin:1998tm}. It
involves a 2--form field $B$ whose equation of motion is required to be (in our convention) $d(dB + C_3) = dC_3.$ This requirement gives rise to the nonlinear self--duality condition for M5--brane. The action has kappa--symmetry provided that self--duality condition is imposed. After appropriate scalings, the on--shell action of \cite{Sezgin:1998tm} is exactly the same as the action \eq{Sfull-onshell}. Explicitly, the identification is $C_3 = -C_3', C_6 = -2C_6', S_{\textrm{M}5}^{\textrm{on--shell}}= 2S'^{\textrm{on--shell}}_{\textrm{M}5},$ where primed quantities are the quantities from \cite{Sezgin:1998tm}.

It is also possible to compare with various M5--brane actions which use PST formalism. For example, it can easily be seen that the action \eq{Sfull-onshell} is the same as the on--shell action of \cite{Pasti:1997gx}, \cite{Bandos:1997ui}, \cite{Ko:2013dka}, \cite{Ko:2017tgo} provided that the actions are scaled appropriately.

The agreement of the on--shell actions implies the agreement of some properties of solitonic solutions. For example, the value of the on--shell action determines the tension of a self--dual string solution.

\setcounter{equation}0
\section{On the form of $R$}\label{sec:solveR}
We have seen that the action \eq{nonlinSenAction} which describes nonlinear self--dual 3--form field (as well as the action \eq{Sfull} which describes M5--brane) possesses all the required symmetries. Furthermore, it is interesting that in order to obtain the symmetries, we do not need to know the form of $R$ as a function of $Q, J, g.$
Nevertheless, in order to understand more about the action \eq{nonlinSenAction}, it could be useful to be able to explicitly write the form of $R.$ Let us give an initial investigation along this direction.

In this section, we restrict ourselves to theories which describe a $*$--self--dual 3--form. The $*$--self--duality is either linear or nonlinear. In the latter case, we do not restrict to just M5--brane.

\subsection{Linear self--dual case}\label{subsec:linsd}
Let us first consider the linear self--dual case.
The task to determine $R$
is equivalent to the task
to determine $\cM$ from the equation
\be
H = Q-R,\qquad
R = \cM Q,
\ee
subject to $*$--self--duality of $H,$ $*'$--self--duality of $Q,$ and $*'$--anti--self--duality of $R.$ Furthermore, $\cM$ is assumed to be symmetric in the sense that $\o_1\w\cM \o_2 = \o_2\w\cM \o_1$ for any $*'$--self--dual 3--forms $\o_1$ and $\o_2.$

There have been two approaches to derive $\cM.$ The first one \cite{Sen:2015nph}, \cite{Sen:2019qit} is to essentially write down $\cM$ formally using the help of some inverse operator.
The second one \cite{Andriolo:2020ykk} is to first choose the basis and write $\cM$ in that basis.
For our purpose, we find it more convenient
to derive a form of $\cM$ alternative to these two approaches.
Let us present our way to obtain this.

The quantity $\cM$ can be thought of as a map from
a $*'$--self--dual 3--form to a $*'$--anti--self--dual 3--form.
The requirements from self--duality and anti--self--duality
are
\be\label{cMoriginalcond}
*'\cM\o_+ = -\cM\o_+,\qquad
(1-*)(1-\cM)\o_+=0,
\ee
where $\o_\pm \equiv (1\pm *')\o/2$ for any 3--form $\o.$
It would be convenient to extend the domain of $\cM$ to a set of any 3--form.
In principle, there are many possible ways to do this.
A simple example way is given in \cite{Andriolo:2020ykk},
in which $\cM$ is chosen such that $\cM\o_- = 0$
and that $(1-*)(1+*' - 2\cM) = 0.$
For us, however, we choose an extension in which $\cM$
satisfies
\be\label{spcMsp}
*'\cM*' = -\cM,
\ee
\be\label{scMsp}
**' - *\cM*' = 1-\cM.
\ee
This clearly differs from the choice chosen in \cite{Andriolo:2020ykk}.
Nevertheless, it is possible to show that the choice of $\cM$ that
we have chosen indeed satisfies \eq{cMoriginalcond} and is also symmetric. The check that our $\cM$ satisfies eq.\eq{cMoriginalcond} is straight forward.
As for the check of symmetry, let us approach as follows.

We first note that the solution to eq.\eq{spcMsp}--\eq{scMsp}
is given by
\be\label{cMSen}
\cM = (1+**')^{-1}(1-**').
\ee
This form is in a similar sense as that given in \cite{Sen:2015nph}, \cite{Sen:2019qit}
since it involves inverse operator.
Note however that eq.\eq{cMSen}
is not simply a formal expression.
The operator $**'$ is a map from 3--forms to 3--forms.
So it can be represented by a square matrix.
With the help of Cayley--Hamilton formula,
the expression $(1+**')^{-1}$ should in fact
be given as a power series in $**'$ up to $19$th order.
The coefficients involve expressions of the form $\tr((**')^n)$
for non--negative integers $n.$
It is complicated to write it in this way and does not provide useful insights for our purpose.
Instead, it is more convenient to write eq.\eq{cMSen} as an infinite power series
\be
\cM
=\sum_{n=1}^\infty (-1)^n((**')^n - (*'*)^n).
\ee
Then by noting that
\be
\o_1\w(**')^n \o_2 = -\o_2\w(*'*)^n \o_1,
\ee
for any 3--forms $\o_1, \o_2$ and any non-negative integer $n,$
it is easy to see that $\cM$ is symmetric.

It can easily be seen that when restricted to flat spacetime, we have $\cM = 0$ as expected.

\subsection{Nonlinear self--dual case}\label{subsec:nonlinsd}
Let us now turn to nonlinear self--dual case.
Consider nonlinear $*$--self--duality condition \eq{nonlin-self-dual-cond} with $\cV$ given by eq.\eq{cV-nonlin}.
We first need to make sure that these equations are consistent after we express
$H^J = Q-R+J$ with $Q = *'Q$ and $R = -*'R.$
This would give a relationship between $F_1([(K^J)^2])$ and $F_2([(K^J)^2]),$
which in turn restrict the form of $U.$

Let us first discuss the flat case.
Imposing an ansatz
\be\label{ansatzR}
R^J = G([(K^J_Q)^2]) (Q^J)^3,
\ee
where $(K^J_Q)_\m{}^\l
\equiv (Q^J)_{\m\n\r}(Q^J)^{\l\n\r},$
and substituting into eq.\eq{cV-nonlin} gives
\be\label{cV-nonlin-flat}
\begin{split}
Q^J + G(Q^J)^3
&=\lrbrk{F_1 -  F_2 G [(K_Q^J)^2] \lrbrk{\hlf + \ove{36}G^2 [(K_Q^J)^2]}}Q^J\\
&\qquad+\lrbrk{ - F_1 G + F_2 \lrbrk{1 + \hlf G^2 [(K_Q^J)^2]}}(Q^J)^3.
\end{split}
\ee
Note that $Q^J$ is $*'$--self--dual but $(Q^J)^3$
is $*'$--anti--self--dual.
So separately considering self--dual and anti--self--dual part of eq.\eq{cV-nonlin-flat} gives
\be\label{Rcondflat-1}
1 = F_1 -  F_2 G [(K_Q^J)^2] \lrbrk{\hlf + \ove{36}G^2 [(K_Q^J)^2]},
\ee
\be\label{Rcondflat-2}
G =  - F_1 G + F_2 \lrbrk{1 + \hlf G^2 [(K_Q^J)^2]}.
\ee
Furthermore, by substituting the ansatz \eq{ansatzR} into $[(K^J)^2],$ we obtain
\be\label{Rcondflat-3}
[(K^J)^2]
=\lrbrk{1+G^2[(K^J_Q)^2]+\ove{36}G^4[(K^J_Q)^2]^2}[(K^J_Q)^2].
\ee
Eliminating $G$ and $[(K^J_Q)^2]$ from eq.\eq{Rcondflat-1}--\eq{Rcondflat-3} gives
\be\label{F1F2-flat}
18 F_1^2 - 18 F_1^4 + 9F_1^2 F_2^2 [(K^J)^2]
-F_2^4[(K^J)^2]^2 = 0.
\ee

The condition \eq{F1F2-flat} relates $F_1$ and $F_2.$
It can be seen that $F_2=0$ corresponds to linear $*'$--self--duality condition.
By setting $F_2=0$ in the condition \eq{F1F2-flat}, we obtain $F_1 = 0, 1, -1.$ However, only $F_1 = 1$ satisfies the original equations \eq{Rcondflat-1}--\eq{Rcondflat-3}.
So we expect that setting $F_2\neq 0$
in the condition \eq{F1F2-flat} would correspond to nonlinear self--duality condition provided that we only choose the solutions which reduce to 
linear self--duality condition in the weak field limit as these solutions would satisfy the original equations \eq{Rcondflat-1}--\eq{Rcondflat-3}.

An example of nonlinear self--duality condition which satisfies eq.\eq{F1F2-flat} and reduces to linear self--duality condition in the weak field limit
is the condition \eq{cV-M5} which is inspired from M5--brane. In this case, the form of $R$ can be obtained and is given by
\be
R^J=\sum_{n=0}^\infty\frac{(-1)^{n+1}}{3n+2}\binom{4n+1}{n}\ove{2^{6n+2}}\frac{[(K_Q^J)^2]^n}{6^n}(Q^J)^3.
\ee
This result can in turn be used to write down eq.\eq{Sfull}
explicitly in terms of $P, Q, g, C_3, C_6.$

It would also be interesting to consider eq.\eq{F1F2-flat} perturbatively in $[(K^J)^2].$
Let us assume that $F_1$ and $F_2$ can be expanded as
\be
F_1([(K^J)^2])
=F_1(0) + [(K^J)^2] F_1'(0) + \hlf[(K^J)^2]^2 F_1''(0) + \cdots,
\ee
\be
F_2([(K^J)^2])
=F_2(0) + [(K^J)^2] F_2'(0) + \hlf[(K^J)^2]^2 F_2''(0) + \cdots.
\ee
Then the leading order of eq.\eq{F1F2-flat}
gives $F_1(0) = 0,\pm 1.$
Again, let us only consider $F_1(0) = 1,$ since this is the only case which reduces to linear self--duality condition in the weak field limit. So after proceeding to solve eq.\eq{F1F2-flat} perturbatively, we obtain
\be\label{F1flat}
F_1([(K^J)^2])
= 1 + \frac{[(K^J)^2]}{4} F_2^2(0) + \hlf[(K^J)^2]^2 \lrbrk{-\frac{17}{144}F_2^4(0) + F_2(0) F_2'(0)} + \cdots,
\ee
\be\label{F2flat}
F_2([(K^J)^2])
=F_2(0) + [(K^J)^2] F_2'(0) + \hlf[(K^J)^2]^2 F_2''(0) + \cdots.
\ee
Substituting eq.\eq{F1flat}--\eq{F2flat} into eq.\eq{Rcondflat-1}--\eq{Rcondflat-3} gives
\be
\begin{split}
G( [(K_Q^J)^2]) &= \frac{F_2(0)}{2}+\frac{1}{2}  [(K_Q^J)^2] F'_2(0)\\
&\qquad+\frac{[(K_Q^J)^2]^2 \left(-F_2^5(0)+144 F_2^2(0) F'_2(0)+288 F''_2(0)\right)}{1152} + \cdots.
\end{split}
\ee

The expressions \eq{F1flat}--\eq{F2flat} can also be substituted into eq.\eq{UfromF1F2} which gives
\be\label{Uflat-universal}
U=\hlf [(K^J)^2] F_2(0) + \ove{8} [(K^J)^2]^2(-F_2^3(0) + 6 F_2'(0)) + \cdots.
\ee
So theories for nonlinear self--dual 3--form in six dimensional flat spacetime should possess $U$ which takes the universal form \eq{Uflat-universal}.

At this point, let us compare this feature of universality to the case of PST formalism.
In PST formalism, an action describing a 2--form field with nonlinear self--dual field strength can also be constructed. This is with the help of auxiliary fields.
Actions in this formalism possess PST1 and PST2 symmetry.
The presence of PST1 symmetry ensures that nonlinear self--duality condition
can be obtained as an equation of motion.
As for PST2 symmetry, its presence ensures
that auxiliary fields indeed contains no dynamics.
In order to see what kind of nonlinear theories are possible, one starts from writing down an action
which contains an undetermined potential term.
The form of this potential term would correspond to the form of nonlinear self--duality condition.
Requiring that the action possesses PST2 symmetry would give rise to
the condition called the PST consistency condition which is imposed on the form of the potential term.
This condition can then be solved which gives rise to the universal form of the potential term \cite{Buratti:2019guq}.
Note that the PST consistency condition is related to the diffeomorphism invariant (or Lorentz invariant in the case of flat spacetime) requirement in the following sense. After gauge fixing the auxiliary fields, the theory would be non manifestly diffeomorphism invariant. The gauge fixed version of the PST consistency condition ensures that the theory is indeed diffeomorphism invariant.

It can be seen that the situation in Sen formalism is analogous to this provided that we realise $U$ as a potential term. Conditions which restrict the form of $U$ comes from requiring the nonlinear theory to be invariant under the nonstandard Lorentz invariant and from the relationship between the physical field $H$ and the unphysical $*'$--self--dual field $Q.$

Let us now turn to discuss Sen formalism in curved spacetime.
We expect that in this case, due to nonstandard transformation of $Q^J$
under diffeomorphism,
promoting equations \eq{Rcondflat-1}--\eq{Rcondflat-3} to include
the metrics would be very complicated
since they depend explicitly on $Q^J.$
So instead of directly attempting to
promote these conditions, let us turn to
the condition \eq{F1F2-flat}.
This condition has $Q^J$ appeared only implicitly
through the field $H^J,$ which transforms in the standard way
under both types of Lorentz transformations.
So eq.\eq{F1F2-flat} could then be directly promoted
to the case of curved spacetime.
It is then easy to see that the promoted condition is covariant.

Since eq.\eq{F1F2-flat} can be promoted to the case of curved spacetime, results analysed in the flat case also hold in the curved case. That is, the linear self--duality condition and the M5--brane nonlinear self--dual condition are consistent. Furthermore, the potential $U$ for consistent nonlinear self--dual theories should be perturbatively expressible as in eq.\eq{Uflat-universal}.

However, when considering the form of $R,$ the extension is not so trivial. That is, in the curved case $R^J$ would no longer be of the form \eq{ansatzR}. In order to get some insights, let us consider perturbative expansion.
Let us start from
\be
\cV = H^J + \a_1(H^J)^3 + \b_1[(K^J)^2]H^J + \a_2 [(K^J)^2](H^J)^3 + \b_2 [(K^J)^2]^2 H^J + \cdots,
\ee
where $\a_1, \a_2, \b_1, \b_2,\cdots$ are functions of $F_2(0), F'_2(0), F''_2(0), \cdots.$
The coefficients can be worked out from eq.\eq{F1flat}--\eq{F2flat}. However, for the following discussion
we do not need explicit forms of these coefficients.
Next, by writing
\be
R^J = R^J_1 + R^J_3 + R^J_5 + \cdots,
\ee
we may obtain the expressions at each order
as follows.
At the leading order, we have
\be
R^J_1 = \cM Q^J.
\ee
At the third order, we have
\be
(1+**')R_3^J
= \a_1((1-\cM)Q^J)^3,
\ee
where we have used $*'$--anti--self--duality of $R_3^J.$
This gives
\be
R^J_3 = \a_1\frac{\cM+1}{2}((1-\cM)Q^J)^3.
\ee
At the fifth order, we obtain
\be
R^J_5 = \b_1 [(K^J_1)^2]\lrbrk{\frac{\cM+1}{2}}(1-\cM)Q^J - 3\a_1\lrbrk{\frac{\cM+1}{2}}\left\{((1-\cM)Q^J)^2 R_3^J\right\},
\ee
where $(K^J_1)_\m{}^\l
\equiv ((1-\cM)Q^J)_{\m\n\r}((1-\cM)Q^J)^{\l\n\r},$
and
\be
\left\{A^2 B\right\}_{\m\n\r}\equiv
\ove{3}
\lrbrk{A_{[\m|\n'\r'}A^{\m'\n'\r'}B_{\m'|\n\r]}
+A_{[\m|\n'\r'}B^{\m'\n'\r'}A_{\m'|\n\r]}
+B_{[\m|\n'\r'}A^{\m'\n'\r'}A_{\m'|\n\r]}}.
\ee
The calculation can be done up to any desirable orders.
This allows us to see a general feature that $R^J$ depends on $g$ only through the map $\cM$ and through raising and lowering of indices. This is interesting that the nonlinearisation does not introduce extra kinds of dependency on $g.$

\setcounter{equation}0
\section{Conclusion}\label{sec:conclusion}
In this paper, we have extended the action in Sen formalism \cite{Sen:2015nph}, \cite{Sen:2019qit} to a complete M5--brane action in Green--Schwarz formalism.
The action is shown to have all the required symmetries.

The action from Sen formalism which is relevant to us is an action of a 2--form field $P$ and a 3--form field $Q,$ both of which transform in a non--standard way under diffeomorphism transformation. The action is given in such a way that a certain combination $H$ of metric and $Q$ satisfies
a self--duality condition off--shell and that $H$ is closed on--shell.
A 3--form source $J$ can also be introduced into the theory.

We have shown by explicit construction that it is possible to extend to theories for nonlinear self--dual 3--form field.
The theories have all the required gauge symmetries as well as diffeomorphism symmetry. It turns out remarkably that in order to prove the symmetries, one only needs to know
the properties of the fields as well as the form of the action \eq{nonlinSenAction} where $U$ is a function of $H^J(Q,g,J), g.$ The explicit form of $H^J(Q,g,J)$ is not needed to be expressed. Furthermore, the symmetries are satisfied as long as the conditions \eq{cond-nonlin-1} and \eq{cond-nonlin-3} are met.
We expect that the analysis in section \ref{sec:nonlin} can easily be extended to describe nonlinear action for chiral $2n-$form in $4n+2$ dimensions in Sen formalism.

We then focus on the action with an appropriate form of $U = U_{\textrm{M}5}.$ After including a 6--form gauge field,
this action can be extended to the action \eq{Sfull} which fully describes an M5--brane action in an eleven--dimensional supergravity background.
The prove of gauge symmetries and kappa--symmetry can be done without the need to explicitly express $H^J(Q,g,J).$
We have also compared the action that we have obtained with some alternative forms of the action in the literature. By appropriate scaling and changing to the same conventions, the actions agree on--shell.
The agreement suggests that they provide the same physics of solitonic solutions.

In the analysis explained so far
one does not need the explicit form of the action in terms of $Q, g, J.$
Nevertheless, in principle it is
important to be able to at least see whether
this can consistently be done.
In the linear case, as shown in 
\cite{Sen:2015nph}, \cite{Sen:2019qit}, \cite{Andriolo:2020ykk},
and further confirmed in subsection \ref{subsec:linsd},
the explicit form of the action can be given.
In the flat nonlinear case,
in order to ensure the consistency,
the conditions \eq{Rcondflat-1}--\eq{Rcondflat-3}
should be satisfied.
These conditions ultimately restrict the form
of the nonlinear action.
It turns out that
the nonlinear self--dual theory
obtained from M5--brane indeed satisfy these conditions.
As for other nonlinear self--dual theories,
we demonstrate a universal perturbative form
of the potential $U.$ This universality is analogous
to the analysis \cite{Buratti:2019guq} in PST formalism.

Note that while the conditions
\eq{Rcondflat-1}--\eq{Rcondflat-3}
do not transform in the standard way under standard Lorentz transformation,
these conditions give rise to the condition \eq{F1F2-flat}
which does transform in the standard way.
The condition \eq{F1F2-flat} can also be thought of as the condition which restricts the form of the nonlinear self--duality condition
provided that we choose only
the theories which reduce to linear self--dual theory
in the weak field limit.
When turned to the curved nonlinear case,
we expect that promoting eq.\eq{Rcondflat-1}--\eq{Rcondflat-3}
would be complicated.
On the other hand, promoting eq.\eq{F1F2-flat}
would be straightforward.
This promoted equation can then be used as the condition
which restricts the form of the action.
So the results at this part would be similar to the flat case,
namely that the M5--brane nonlinear self--dual theory
is allowed and that the potential $U$
on any allowed theories should take the universal
perturbative form as presented in eq.\eq{Uflat-universal}.

In order to write the action explicitly in terms of
$Q, g, J,$ one needs to be able to determine $R.$
For the case of flat nonlinear theories,
we have solved the conditions \eq{Rcondflat-1}--\eq{Rcondflat-3}
to obtain the explicit form
for M5--brane nonlinear self--dual theory
and the perturbative expression
for other consistent nonlinear self--dual theories.
As for the curved nonlinear theories,
although we have shown that M5--brane theory is consistent,
it would still be interesting to write the action
explicitly in terms of $P, Q, g, C_3, C_6.$
We have made an initial attempt along this direction by making perturbative analysis.
It turn out that the appearances of the metric $g$
in nonlinear theories are only in raising and lowering
indices as well as through the quantity $\cM$
in the form we present in eq.\eq{cMSen}.
We expect that in order to be able to provide the full analysis,
the conditions \eq{Rcondflat-1}--\eq{Rcondflat-3}
should be promoted to include the metric $g.$
This is a highly nontrivial task, so leave this as a future work.

The explicit form of the nonlinear action
in terms of $Q, g, J$ would be required
in order to perform the analysis of dimensional reduction,
which will provide further physical insights
to the theories.
In fact, although the explicit full form of nonlinear theory
in terms of $Q, g, J$ in curved spacetime is not yet known,
dimensional reduction can still be analysed.
This is because in this kind of analysis
one specifies the form of the metric.
So although the metric appears in quite
a complicated way in the full theory,
dimensional reduction could be relatively simpler.
As a future work, we will most likely proceed along this direction to
see what theories would arise from the dimensional reductions.

For the analysis in this paper, we assume that the six--dimensional spacetime has a trivial topology. This is to ensure that being closed on--shell, $H$ is exact. It would be interesting to extend the analysis to the case of nontrivial topology and to see if it is possible to compare and contrast with PST formalism in spacetime with nontrivial topology \cite{Bandos:2014bva}, \cite{Isono:2014bsa}.

Finally, there is also another interesting recent construction \cite{Mkrtchyan:2019opf} of chiral 2--form theory. This construction is equivalent to PST formalism. So it could be possible to also obtain a complete M5--brane action based on this construction.

\section*{Acknowledgements}
The author would like to thank Dmitri Sorokin and Sheng-Lan Ko
for discussions and very helpful comments on the
manuscript.

\providecommand{\href}[2]{#2}\begingroup\raggedright\endgroup

\end{document}